\documentclass[useAMS,usenatbib]{mn2e}

\usepackage{graphicx}
\usepackage{mathabx}

\title[The case for primordial black holes as dark matter]{The case for
 primordial black holes as dark matter}
\author[M. R. S. Hawkins]{M. R. S. Hawkins$^{1}$\thanks{E-mail:
 mrsh@roe.ac.uk}\\
 $^{1}$Institute for Astronomy (IfA), University of Edinburgh,
 Royal Observatory, Blackford Hill, Edinburgh EH9 3HJ, UK}
\begin{document}

\date{Accepted 1988 December 15. Received 1988 December 14; in original
form 1988 October 11}

\pagerange{\pageref{firstpage}--\pageref{lastpage}} \pubyear{2011}

\maketitle

\label{firstpage}

\begin{abstract}

The aim of this paper is to present the case that stellar mass primordial
black holes make up the dark matter component of the Universe.  A near
critical density of compact bodies implies that most lines of sight will
be gravitationally microlensed, and the paper focuses on looking for
the predicted effects on quasar brightness and spectral variations.
These signatures of microlensing include the shape of the Fourier power
spectrum of the light curves, near achromatic and statistically symmetric
variations, and the absence of time dilation in the timescale of
variability.  For spectral changes it is predicted that as the continuum
varies there is little corresponding change in the strength of the broad
lines.  In all these cases, the observations are found to be consistent
with the predictions for microlensing by a population of stellar mass 
compact bodies.  For multiply lensed quasar systems where the images vary
independently and microlensing is the generally accepted explanation,
the case is made that stellar populations are too small to produce the
observed effects, and that the only plausible alternative is a
population of compact dark matter bodies of around a stellar mass
along the line of sight.  The most serious objection to dark matter in
the form of compact bodies has come from observations of microlensing of
stars in the Magellanic Clouds.  In this paper the expected event rate
is re-analysed using more recent values for the structure and dynamics
of the Galactic halo, and it is shown that there is then no conflict
with the observations.  Finally, the possible identity of a near
critical density of dark matter in the form of stellar mass compact
bodies is reviewed, with the conclusion that by far the most plausible
candidates are primordial black holes formed during the QCD epoch.  The
overall conclusion of the paper is that primordial black holes should be
seen alongside elementary particles as viable dark matter candidates.

\end{abstract}

\begin{keywords}
dark matter -- cosmology: observations
\end{keywords}

\section{Introduction}
\label{sec1}

The nature of dark matter remains one of the biggest unsolved problems in
cosmology.  Over the last 20 years or so, the identification of dark matter
with elementary particles has received widespread support.  This idea is
primarily motivated by the acceptance that the observed density of dark
matter far exceeds the density of baryonic material predicted to be created
in the standard Big Bang cosmology, with the implication that the dark
matter must be in non-baryonic form.  Elementary particles have been seen
as the most obvious choice here as, despite the lack of any observational
evidence, theoretical predictions from extensions to the standard model of
particle physics provide a number of different particles as dark matter
candidates.  Much effort has been put into detecting such particles
directly, but so far there has been no success \cite{f10a}.  In fact the
parameter space to be searched is very large, and potential dark matter
particles have by no means been ruled out, but it is perhaps worthwhile at
this stage to review the status of other candidates.

\begin{figure*}
\centering
\begin{picture} (0,280) (250,0)
\includegraphics[width=1.0\textwidth]{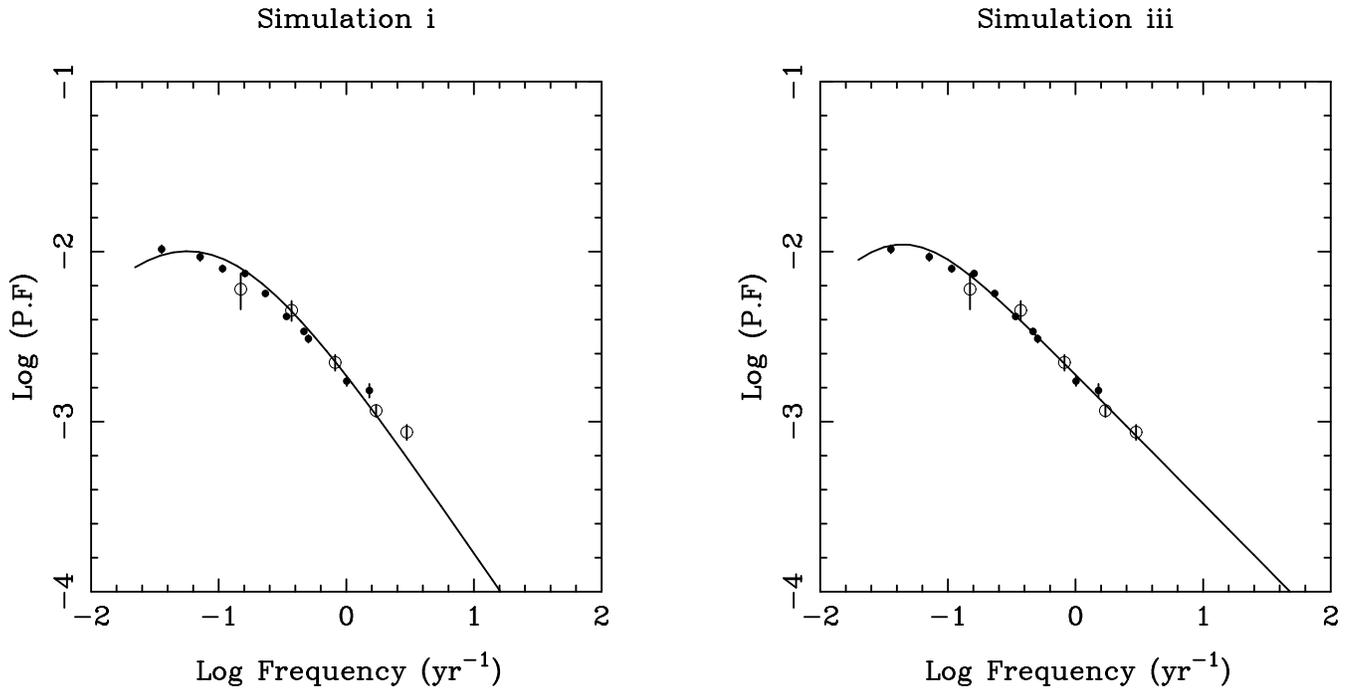}
\end{picture}
\caption{SEDs of light curves in the observer frame for quasars from the
 Field 287 survey (filled circles) and from the MACHO project (open
 circles).  The solid lines are SEDs for simulated light curves from Minty
 (2001). In the left hand panel $\Omega_{Lens}=1.0$ and
 $\Omega_{\Lambda}=0.0$.  In the right hand panel $\Omega_{Lens}=0.3$ and
 $\Omega_{\Lambda}=0.7$.}
\label{fig1}
\end{figure*}

The requirement that dark matter be non-baryonic puts a major constraint
on any alternative to elementary particles.  In particular, the idea that
dark matter is in the form of compact bodies has been treated with some
scepticism, as familiar objects such as stellar remnants or brown dwarfs,
being baryonic, are ruled out as candidates.  However, there are
possibilities for compact bodies which are non-baryonic.  The most
thoroughly investigated of these are primordial black holes.  A strong case
has been made that in the early Universe density fluctuations could lead to
copious production of primordial black holes \cite{c74,c75}.  Such black
holes are most likely to be formed at epochs when the Universe undergoes
phase transitions, with masses dominated by the horizon mass at formation
\cite{j99}.  In particular, black holes formed during the QCD phase
transition a few microseconds after the Big Bang should have masses peaking
at the QCD horizon mass scale of around $1 M_{\odot}$ \cite{j97}.
As these objects would be created well before the epoch of baryon
synthesis, they would not be subject to the constraints on baryon mass
density.

The detection of dark matter in the form of compact bodies has received
much attention in the literature.  In an early paper, Press \& Gunn (1973)
pointed out that in a universe where the density of matter is equal to the
critical density, every line of sight will be gravitationally lensed.
This means that if the dark matter is in the form of compact bodies, they
will have the effect of distorting the image of any distant light source.
For bodies of around a solar mass the characteristic
timescale for changes in brightness is a few years.  This domain of
gravitational lensing, known as microlensing, opens up the potential for
detecting dark matter in this form by observing changes in brightness of
distant compact light sources such as quasars.  Much work has been done by
several groups in simulating the light curves expected from microlensing,
with the idea of comparing the results with observed quasar variability
\cite{c79,p86a,k86,s87}.  There is however a problem with this approach, as
any intrinsic quasar variability must be taken into account. 

There is one situation where quasar microlensing can be identified
unambiguously.  There are now many known examples of a massive galaxy
producing multiple images by the gravitational lensing of a more distant
quasar along the line of sight \cite{w79,h85}.  Any intrinsic variation in
the quasar will be observed in all images, with some time difference due to
the differing path lengths to the images.  However, in all quasar systems
which have been adequately observed, it has been found that for the most
part individual images vary independently.  This is widely recognised as
being due to microlensing \cite{i89}, and supports the idea that all
quasars are being microlensed.  There is however a caveat, that as these
systems are known to have a massive galaxy close to the line of sight,
normal stars in this lensing galaxy might be responsible for the
microlensing.

\begin{figure*}
\centering
\begin{picture} (0,280) (250,0)
\includegraphics[width=1.0\textwidth]{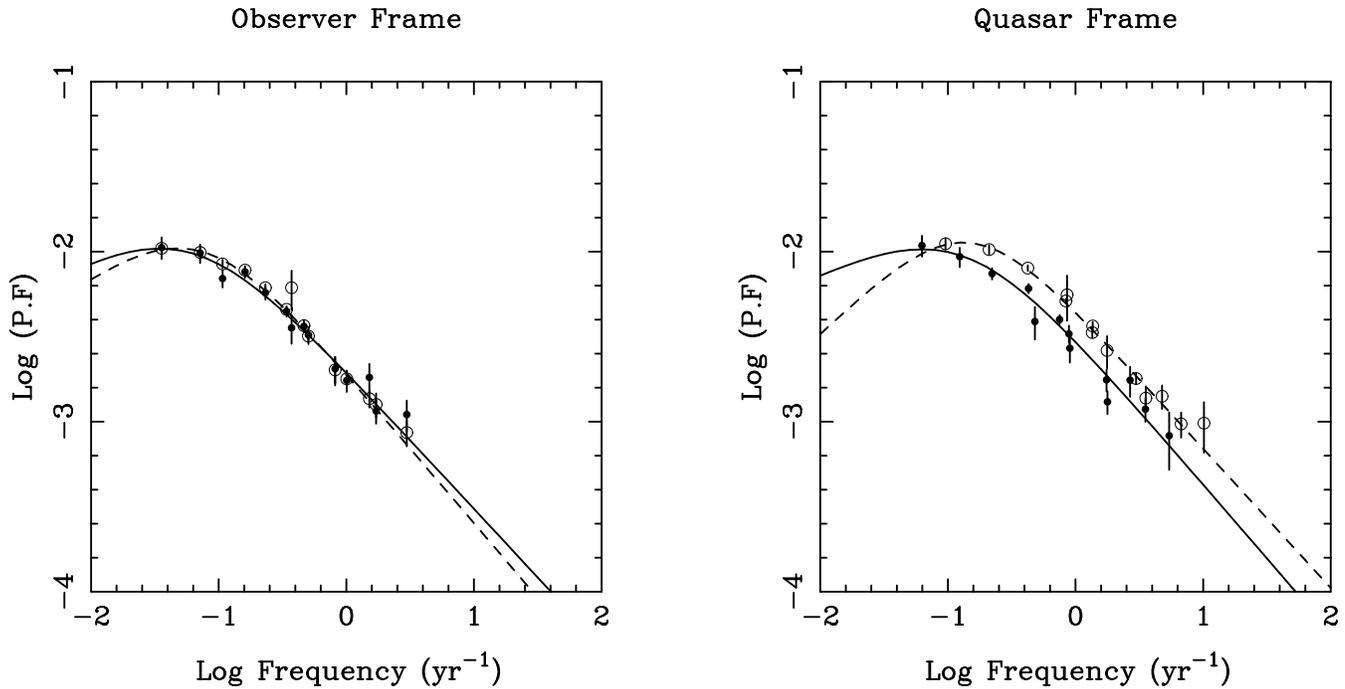}
\end{picture}
\caption{The left hand panel shows the superposition of the SEDs for
 low redshift (filled cicles) and high redshift (open circles) samples of
 light curves.  The solid and dashed lines are best fits of the curve in
 Eq.~\ref{eqn7}.  The right hand panel shows SEDs of the same light curves
 in the quasar frame, with a correction for time dilation applied.} 
\label{fig2}
\end{figure*}

Perhaps the best known search for compact bodies is for those in the halo
of the Galaxy.  By monitoring several million stars in the LMC and SMC,
bodies of around a solar mass were detected by the MACHO project
\cite{a00a}.  The detection rate of these bodies implies a larger population
than can be accounted for by known stellar populations, and so they are
clearly good dark matter candidates.  However, they have been discounted
on the basis that they make up insufficient dark matter for the
predictions of models of the Galactic halo consistent with available
observations.

The purpose of this paper is to review the arguments that compact bodies in
the form of primordial black holes make up the dark matter component of the
Universe.  We shall firstly look at the evidence for microlensing in quasar
light curves, which includes time dilation measures, comparison with
microlensing simulations, chromatic changes and statistical symmetry.
Much of this work has already been published, but the comparison of the
simulations with observations is new, and work on colour changes and
symmetry is updated.  We derive a further test for microlensing by
comparing changes in continuum and emission line flux.  Although this
has been done before, we use new and very much improved data here.  We then
examine from a new perspective the claim that microlensing in multiply
imaged quasar systems is caused by stars in the lensing galaxy.  Finally,
we use the Galactic halo microlensing observations of the MACHO
collaboration together with the most recent measurements of the halo
parameters to re-determine the optical depth to microlensing, and hence
the limit on dark matter in the form of compact bodies.

\section{Quasar light curves}
\label{sec2}
\subsection{Microlensing simulations}

\begin{figure*}
\centering
\begin{picture} (0,200) (250,0)
\includegraphics[width=1.0\textwidth]{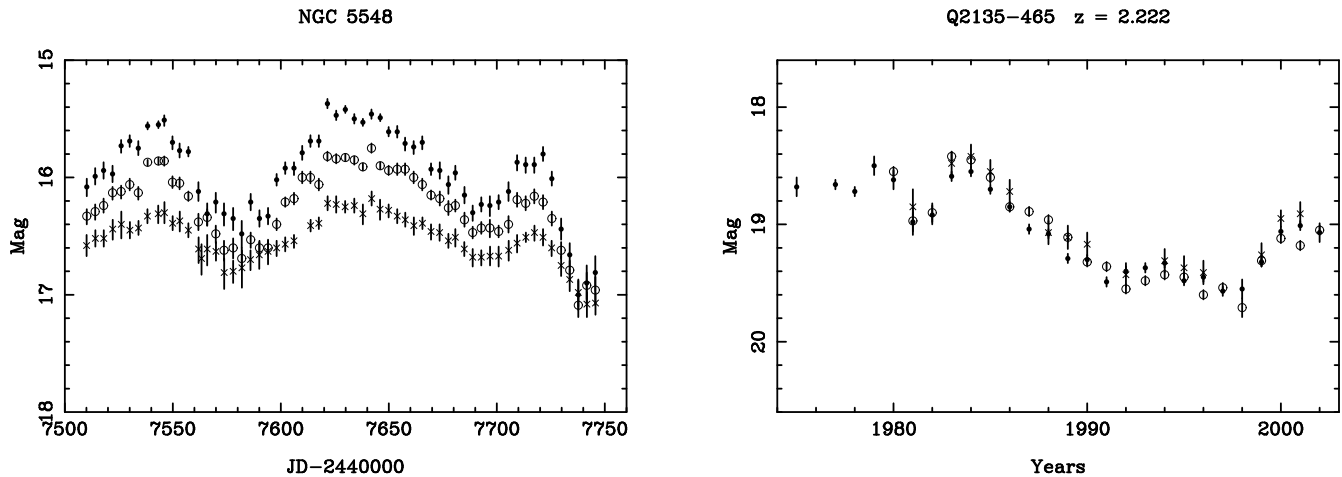}
\end{picture}
\caption{The left hand panel shows continuum ultraviolet light curves for
 the Seyfert galaxy NGC 5548 from the IUE satellite.  The wavelength bands
 are centred on $\lambda 1350$ (crosses), $\lambda 1840$ (filled
 circles) and $\lambda 2670$ (open circles).   The right hand panel shows
 optical light curves for a high redshift quasar from the Field 287
 survey.  The wavelength bands are $U$ (crosses), $B_{J}$ (filled
 circles) and $R$ (open circles).}
\label{fig3}
\end{figure*}

If primordial black holes make up the bulk of dark matter, then they should
betray their presence by microlensing the light of distant quasars.  The
structure of the resulting light curves has been the subject of extensive
research, in the form of computer simulations and theoretical studies.
Fourier power spectrum analysis is perhaps the most useful tool for
comparing the observed variability of quasars with the results from
numerical simulations, but structure functions and auto-correlation
functions have also been used.

We start with some definitions and formulae.  The Einstein radius
$\theta_{e}$ of a gravitational lens is defined as

\begin{equation}
 \theta_{e} = \left( \frac{4Gm}{c^{2}R} \right)^{1/2}
\label{eqn1}
\end{equation}

\noindent
which thus varies as the square root of the lens mass $m$.  $R$ is a
distance parameter defined by

\begin{equation}
 R = \frac{D_{ol}D_{os}}{D_{ls}}
\label{eqn2}
\end{equation}

\noindent
with the subscripts indicating angular diameter distances between observer,
source and lens; $G$ is the gravitational constant and $c$ is the speed of
light.  For a solar mass primordial black hole midway to a quasar at a
redshift of 1 to 2, $\theta_{e}$ subtends a distance of about $10^{-2}$ pc
or $3 \times 10^{16}$ cm.
For a single isolated lensing event where the source is small compared
with the Einstein radius of the lens, the amplification $a$ is geven by
\cite{v83}

\begin{equation}
 a = \frac{1+2\theta^{-2}}{(1+4\theta^{-2})^{1/2}}
\label{eqn3}
\end{equation}

\noindent
where $\theta$ is the angular distance from the observer's line of sight
to the lens in units of the Einstein radius.  In the case of microlensing,
the change of brightness with time $t$ is thus given by

\begin{eqnarray}
 a(t) = \frac{1+2\theta(t)^{-2}}{(1+4\theta(t)^{-2})^{1/2}}, \nonumber \\
 \nonumber \\
 \theta(t)^{2} = \theta_{0}^{2}+v^{2}(t-t_{0})^{2}
\label{eqn4}
\end{eqnarray}

\noindent
where $\theta_{0}$ is the minimum distance from the trajectory of the
lens to the source in units of Einstein radius, $t_{0}$ is the time of
closest approach of the lens to the source, and $v$ is the speed of
the source in units of Einstein radius per unit time.  For multiple
lenses there are no simple analytic solutions, and the changes in
brightness must be obtained from simulations.

The first simulations of microlensing at large optical depth were published
by Paczy\'{n}ski (1986a).  This early work modelled a single plane with
randomly placed lenses, and a point source moving across it relative to the
observer.  Kayser et al. (1986) extended this work using an inverse ray
tracing technique, to include the case of extended sources.  Improvements
in numerical techniques \cite{s87,r91,l93} enabled more complicated lens
distributions and source parameters to be modelled.  For the purposes of
comparing the observed quasar light curves with flux variations from
microlensing, we shall use the simulations of Minty (2001).  We choose
these simulations as they are the development of the earlier work cited
above, but incorporating a 3-dimensional lens distribution and,
importantly for the purposes of this paper, the results are presented in
the form of Fourier power spectra of the microlensing variations, which is
essential for comparison with observations.  The ray tracing procedure is
described by Minty et al. (2002), together with an illustration of a
simulated microlensing light curve.

In these simulations
the lenses are distributed on multiple planes, and move independently with
a random velocity drawn from a Gaussian distribution.  Rays are traced back
from the observer and their position in the source plane recorded, which
then enables the magnification pattern to be calculated.  The lenses are
then moved according to their random velocity and the ray-tracing procedure
is repeated, allowing the simulation to evolve over time.  The
3-dimensional lens distribution puts great demands on computer power, and
so the simulations were carried out on a number of parallel platforms at
the Edinburgh Parallel Computing Centre (EPCC). 

The magnification patterns in the observer plane allow the variations in
brightness due to microlensing to be converted to simulated light curves
for sources with specified light profiles.  This procedure was used to
generate large numbers of light curves for which Fourier power spectra were
calculated.  This format is well suited to comparing the characteristics of
microlensing variations with observed light curves, for which Fourier power
spectra can also be calculated.  The results of the ray tracing procedure
depend upon the cosmological model and the proportion of dark matter in the
form of compact bodies, and Minty (2001) treats a variety of different
cases. 

\begin{table}
\caption{Simulation parameters.}
\label{tab1}
\centering
\vspace{5mm}
\begin{tabular}{r r r r r r l r}
\hline\hline
 & & & & & & & \\
 Sim & Lenses & Planes & $M_L$ & $z_s$ &
 $\Omega_M$ & $\Omega_L$ & $\Omega_{\Lambda}$ \\
 & & & & & & & \\
\hline
 & & & & & & & \\
   {\it i} & 100000 &  5 &       1 & 2.0 & 1.0 &  1.0 & 0.0 \\
  {\it ii} & 100000 & 10 &       1 & 2.0 & 1.0 &  1.0 & 0.0 \\
 {\it iii} &  87875 &  5 &       1 & 2.0 & 0.3 &  0.3 & 0.7 \\
  {\it iv} &  23000 &  5 &       1 & 1.0 & 0.3 &  0.3 & 0.7 \\
   {\it v} &  29291 &  5 &       1 & 2.0 & 0.3 &  0.1 & 0.7 \\
  {\it vi} &   2929 &  5 &       1 & 2.0 & 0.3 & 0.01 & 0.7 \\
 {\it vii} &  20000 &  5 & 0.1 - 1 & 2.0 & 0.3 &  0.1 & 0.7 \\
 & & & & & & & \\
\hline
\end{tabular}
\end{table}

Table~\ref{tab1} is adapted from Minty (2001), and gives parameters for
the seven models for which simulations are presented.  Column 1 gives
the simulation identification, columns 2 and 3 the number of lenses and
lens planes.  Column 4 gives the range of lens masses in units of
$M_{\odot}$, and column 5 the redshift of the source.  The last three
columns list the cosmological parameters for the simulations.  $\Omega_M$
and $\Omega_{\Lambda}$ are the conventional contributions to the density
parameter from mass and dark energy respectively, and $\Omega_L$ is
the contribution from lenses alone.

The observations which we shall use for comparison with the simulations
are described in detail in Hawkins (2010), and comprise light curves
taken from two sources.  The first is a large scale yearly monitoring
programme of some 1200 quasars in several passbands covering a total of
28 years \cite{h03,h07}.  The second is from the MACHO project, where
quasars detected as part of the monitoring programme have been identified
\cite{g03}.  The original observations cover timescales from a few days
to several years, and form an important complement to the long term light
curves.  In order to make a quantitative comparison with the simulations,
Fourier power spectra were calculated for the two datasets, as described in
Hawkins (2010).

We define the Fourier power spectrum $P(s)$ as:
 
\begin{eqnarray}
P(s_{i}) = \frac{t}{N} \left( \sum_{j=1,N} m(t_{j}) cos \frac{2 \pi
 i j} {N}\right)^{2} +
 \nonumber \\ \frac{t}{N} \left( \sum_{j=1,N}
 m(t_{j}) sin \frac{2 \pi i j} {N}\right)^{2}
\label{eqn5} 
\end{eqnarray}

\noindent
where $i$ runs over the $N$ equally spaced epochs of observation separated
by time $t$, and $m(t_{j})$ is the magnitude at epoch $t_{j}$.  In
the case of a sample of light curves, the integration for each frequency
continues over all sample members.  We define the SED as a plot of the
product of Fourier power and frequency versus frequency.

In Fig.~\ref{fig1} the SEDs of the observed quasar light curves are shown
in both panels.  The data from the two sources described above are plotted
with different symbols, and it will be seen that there is good agreement
in the area of overlap.  This provides a limit on any systematic
differences between the two samples, and a check on the validity of the
error bars.  The solid lines are the SEDs of simulations of microlensing
light curves from Minty (2001).  The left hand panel is for an
Einstein-de Sitter cosmology with lenses making up the critical density,
and the right hand panel for a non-zero cosmological constant, with
$\Omega_{\Lambda}=0.7$, and all matter in the form of lenses with
$\Omega_{Lens}=0.3$.  The position $f$ of the curves on the frequency axis
is a measure of timescale of variation $t$, and in the simulations depends
on two free parameters.  The first is the Einstein radius which is
proportional to the square root of the lens mass $M_L$, and the second is
the mean transverse velocity $v_t$ of the lenses across the line of sight
to the source.  Hence

\begin{equation}
 t \; \propto \; \frac{\sqrt{M_L}}{v_t} \; \propto \; f^{-1}
\label{eqn6}
\end{equation}

\noindent
The characteristic lens mass may thus be measured by adjusting $t$ to fit
the simulations to the data.  However, it will be seen from
Eq.~\ref{eqn6} that there is a degeneracy between $M_L$ and $v_t$.
There are a number of components to $v_t$, which have been measured with
varying degrees of accuracy.  These include the solar motion relative to
the CMB \cite{k93}, peculiar velocities of galaxy populations \cite{h03a},
and bulk motions of the intervening material and the source
\cite{h04,k08}.  Combining these contributions gives $v_t \approx 1300$
km sec$^{-1}$, with a range from 1000 to 1550 km sec$^{-1}$.
Table~\ref{tab2} shows the result of a least squares iterative fit of the
simulations to the data, varying $M_L$ for different values of $v_t$.
The resulting lens masses lie in the range 0.08 -- 0.23 $M_{\odot}$.  Both
simulations in Fig.~\ref{fig1} have a broadly similar shape to the data,
which is independent of the fitting process.

Although the good agreement between the SEDs of the observed light curves
and the microlensing simulations does not show that quasars are being
microlensed, it does demonstrate that if dark matter is in the form of
stellar mass compact bodies, then the variations expected from
microlensing simulations are consistent with observations of quasar light
curves.

\begin{table}
\caption{Lens masses in units of $M_{\odot}$.}
\label{tab2}
\centering
\vspace{5mm}
\begin{tabular}{l r r r}
\hline\hline
 & & & \\
 $v_t$ (km sec$^{-1}$) & 1000 & 1300 & 1550 \\
 & & & \\
\hline
 & & & \\
 Sim {\it i}   & 0.08 & 0.13 & 0.19 \\
 Sim {\it iii} & 0.10 & 0.16 & 0.23 \\
 & & & \\
\hline
\end{tabular}
\end{table}

\subsection{Time dilation}

The idea that primordial black holes are microlensing the light of quasars
can be tested quite simply by looking for the effects of time dilation in
quasar light curves.  If the observed variations are intrinsic to the
quasar, then whatever their origin, they should show a stretching of
timescale with redshift by a factor $(1+z)$.  If on the other hand the
observed variations are dominated by the effects of microlensing, then no
such change in timescale will be seen.  This is because the variations no
longer originate in the quasar rest frame, but at the redshift of the
lenses.  Due to the curvature of space at high redshift, it turns out that
the probability distribution for the position of lenses is strongly peaked
close to a redshift $z = 0.5$, regardless of the redshift of the source
\cite{t84}.  This result is insensitive to the assumed cosmology, and
means that the effects of time dilation on quasar light curves from
microlensing will be small.  Consequently there will be little difference
in the characteristic timescales of low and high redshift samples.

The task of looking for time dilation in quasar light curves was addressed
in a recent paper \cite{h10} using the data discussed earlier in this
Section.  The light curves were divided into high and low redshift samples,
with the additional constraint of a restricted range in absolute magnitude,
to prevent any confusion with possible luminosity effects.  SEDs for the
light curves from the two redshift samples were calculated as described
above, and are plotted in Fig.~\ref{fig2}.  The idea was to look for a
shift to longer timescales for the high redshift sample relative to the
lower redshift one.  To aid comparison, the two SEDs were fitted with a
double exponential function of the form:

\begin{equation} 
P(f) = \frac{C}{\left(\frac{f}{f_{c}}\right)^{a}
 +\left(\frac{f}{f_{c}}\right)^{-b}}
\label{eqn7}
\end{equation}

\noindent
The left hand panel shows the high and low
redshift SEDs superimposed, and it is clear that no shift is observable.
The right hand panel shows the the same data, but with a correction for
time dilation applied.  This has the effect of preferentially moving the
high redshift sample to shorter timescales.  This gives an idea of the
expected shift between the two curves if time dilation were present.

If primordial black holes are the main constituent of dark matter, then
they should cause large variations in quasar light, and these variations
will not show the effects of time dilation.  The data in Fig.~\ref{fig2}
supports this prediction, and also implies that quasar variation from
microlensing dominates any intrinsic process.  Although this does not
necessarily establish that quasars are being microlensed, alternative
explanations for the observations are not easy to find.

\subsection{Colour changes}
\begin{figure}
\centering
\begin{picture} (0,280) (120,0)
\includegraphics[width=0.5\textwidth]{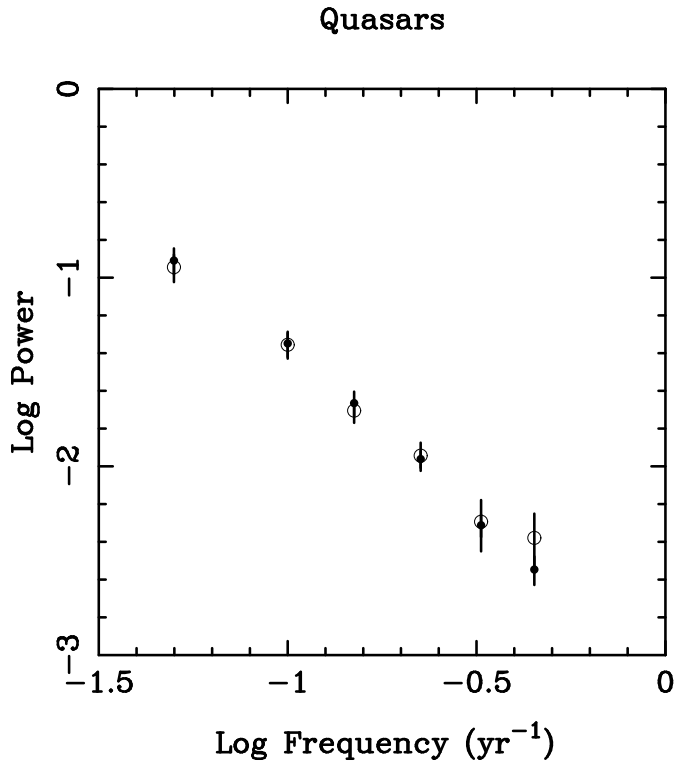}
\end{picture}
\caption{Fourier power spectra for a sample of 75 quasar light curves in
 $B_{J}$ (filled circles) and $R$ (open circles).  The sample limits are
 $z > 1.5$ and $M_{B} < -26$.}
\label{fig4}
\end{figure}
The study of colour changes in AGN has received extensive attention from
campaigns to measure the change in emission line strength relative to
continuum luminosity in Seyfert I galaxies \cite{p02}.  These
observations are designed to measure time delays between nuclear
variations and emission line response, but they also show without any doubt
that the nucleus changes colour in the sense that, as it brightens, it
becomes bluer.  The left hand panel of  Fig.~\ref{fig3} shows light curves
of the nucleus of the Seyfert I galaxy NGC 5548 in three ultra-violet bands
from the IUE satellite \cite{c91}.  The data have been converted to
magnitudes on an arbitrary scale, and show quite clearly the change of
colour with brightness.  The distance to NGC 5548 is much too small for any
variations in brightness to be caused by microlensing, but we now
investigate the situation at higher redshift where gravitational lensing is
a possibility. 

A well known property of gravitational lensing is that for a uniformly
illuminated or point source, the effects of the lensing will be
achromatic as photon trajectories are independent of colour.  If quasars
are being microlensed by primordial black holes, one might therefore
expect that the variations would be achromatic.  This will be the case
provided that quasars are point sources, in the sense that they only
emit light from within an angle that is small compared with the Einstein
radius of the lens.  However, dilution of the nuclear light by light from
the host galaxy, or an extended emission region with a colour gradient
around the nucleus itself, will cause the quasar to change colour as it
brightens or fades.  The procedure for correcting for the effect of the
host galaxy on a quasar light curve has been examined in some detail by
Hawkins (2003). Here we shall confine ourselves to the straightforward
case where the nucleus is sufficiently luminous for the observed
variations not to be significantly affected by the underlying galaxy.

The right hand panel of Fig.~\ref{fig3} shows light curves for a luminous
quasar in the $U$, $B_{J}$ and $R$ bands.  The wavelengths of these bands
in the quasar rest frame are close to the wavelengths of the ultra-violet
bands in the left hand panel, and provide an interesting comparison
between the modes of variation of the two objects.  It is clear that in
contrast to the Seyfert galaxy, the variations of the quasar are
essentially achromatic.

The contrast between the two objects in Fig.~\ref{fig3} may be put on a
statistical footing by comparing the spectrum of variations of a large
sample of red and blue passband quasar light curves.  Fig.~\ref{fig4}
shows Fourier power spectra for 75 such light curves of luminous quasars
($M_{B} < -26$), and it will be seen that the two spectra have the same
power at all frequencies.  This implies that the variations of the sample
as a whole are achromatic.

\subsection{Symmetry}
\begin{figure}
\centering
\begin{picture} (0,280) (120,0)
\includegraphics[width=0.5\textwidth]{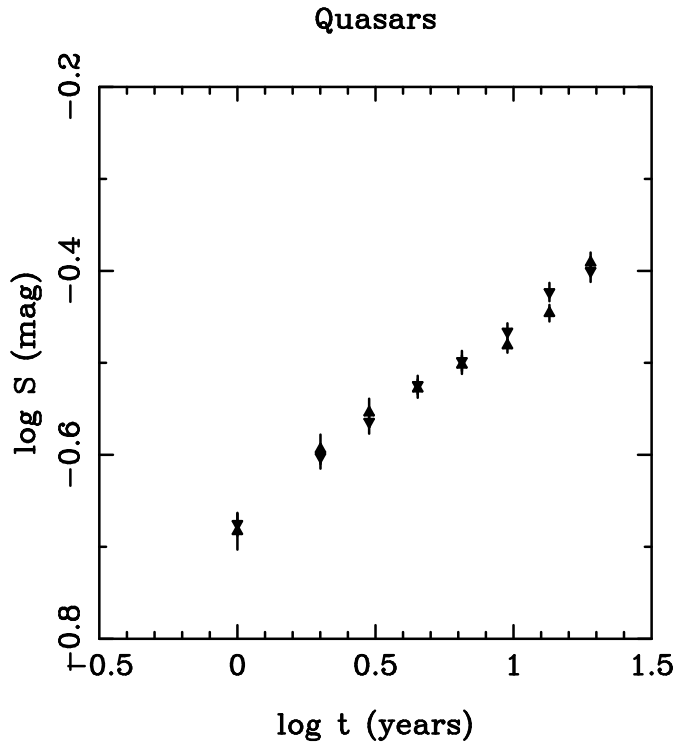}
\end{picture}
\caption{Time asymmetric structure functions for a sample of 366 quasar
 light curves.  Functions for increasing and decreasing brightness are
 shown by upward and downward pointing triangles respectively.  The
 sample limits are $z > 1.5$ and $M_{B} < -23$.}
\label{fig5}
\end{figure}

It is clear from Eq.~\ref{eqn4} that in the restricted case of a single
point source, the brightness variation will be symmetrical in time.
However, in a situation where multiple lenses are involved, the variation
is much more complicated.  The amplification patterns combine in a
non-linear way to produce variations with no obvious symmetrical
structure.  The resulting light curve will however be statistically
symmetric in the sense that there will be no preferred direction of time.
In other words it will not be possible to distinguish between running time
forwards or backwards by examining the variations alone.

The measurement of statistical symmetry in quasar light curves has been
addressed in an interesting paper by Kawaguchi et al. (1998).  They
present an accretion model of quasar variability which is predicted to
lead to asymmetrical variations in the sense that the brightness of the
nucleus will rise more slowly than it falls.  To measure this asymmetry
they define a modification of the structure function $S$ which may be
defined by

\begin{equation}
 S(t) =\sqrt{\frac{1}{N(t)}\sum_{i<j}[m(t_{j})-m(t_{i})]^{2}}
\label{eqn8}
\end{equation}

\noindent
where $m(t_{i})$ is the magnitude measure at epoch $t_{i}$, and the sum
runs over the $N(t)$ epochs for which $t_{j}-t_{i} = t$.  In addition
to the standard structure function defined above, we shall for the purpose
of measuring asymmetries also make use of two modified structure functions
$S_{+}$ and $S_{-}$.  These are defined as for $S$ except that for $S_{+}$
the integration only includes pairs of magnitudes for which the flux
becomes brighter, and $S_{-}$ for which it becomes fainter.

Kawaguchi et al. (1998) calculate the functions  $S_{+}$ and $S_{-}$ for
simulated light curves from their accretion model and show that there is
more power in $S_{-}$ implying a gradual rise and rapid decay in flux.
This is a common feature of accretion models, and a plausible mechanism
for intrinsic variation in quasars.  If dark matter is in the form of
primordial black holes, then the microlensing effect should result in time
symmetric variation as discussed above, with no difference between $S_{+}$
and $S_{-}$.

Fig.~\ref{fig5} shows $S_{+}$ and $S_{-}$ calculated for a sample of quasar
light curves from Field 287.  The two curves are not distinguishable
within the errors of observation, implying that the variations are time
symmetric.  Although this certainly does not rule out the possibility
that the quasar variations are intrinsic, it is as expected for
microlensing by primordial black holes.

\section{Spectral changes in quasars}
\label{sec3}

An important diagnostic for the origin of AGN variability is the analysis
of correlations between emission line and continuum variations.
Reverberation mapping studies of Seyfert I galaxies, notably NGC 5548
\cite{p02}, have conclusively shown that changes in continuum flux are
followed after a few days by a corresponding change in broad emission line
flux.  The way this is generally understood is that the broad emission line
region surrounding the accretion disc and supermassive black hole is
responding to changes in the central photon flux with corresponding
changes in line flux.  The time delay is seen as a measure of the photon
travel time from the central engine to the broad line region and hence a
measure of its size.

For quasar variations caused by microlensing, the changes in spectrum are
predicted to follow a different pattern as a result of the different sizes
of the continuum and broad line emitting regions. The effect of source
size on gravitational lens amplification has been the
subject of a number of investigations.  Refsdal \& Stabell (1991) derived
a simple analytical formula for estimating the typical microlensing
amplification $\delta m$ for sources larger than the Einstein radius of
the lenses, compared to the amplification of point sources:

\begin{equation}
 \delta m \approx \frac{2.17 \sqrt {|\sigma|} \theta_e M}{\theta}
\label{eqn9}
\end{equation}

\noindent
Here, $\sigma$ is the normalised surface density in stars, $\theta_e$
the Einstein radius for the effective mass $M$ of the microlensing stars,
and $\theta$ is the angular radius of the source.  These parameters are
described in more detail by Refsdal \& Stabell (1991).  The important
point to notice is a decrease in amplfication magnitude $\delta m$ for
source size $\theta$ scaling as $(\theta / \theta_{e})^{-1}$, which is
consistent with results from microlensing simulations \cite{k86,s87,l93}.  
\begin{figure}
\centering
\begin{picture} (0,280) (120,0)
\includegraphics[width=0.5\textwidth]{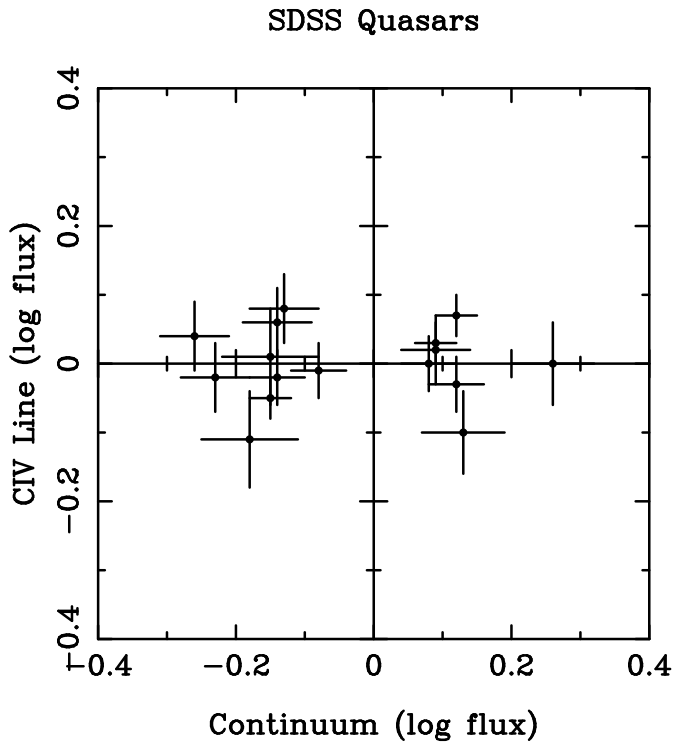}
\end{picture}
\caption{Change in CIV line flux versus change in continuum flux adapted
 from Wilhite et al. (2006)}
\label{fig6}
\end{figure}
\begin{figure}
\centering
\begin{picture} (0,200) (120,0)
\includegraphics[width=0.5\textwidth]{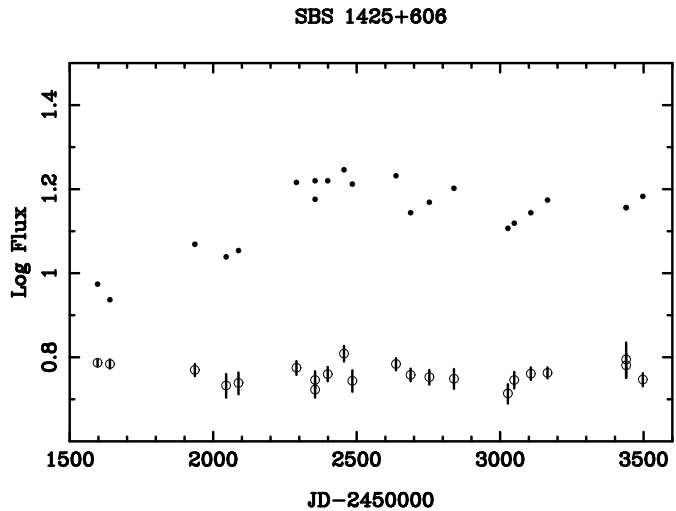}
\end{picture}
\caption{Light curves for the quasar SBS 1425+606 adapted from
 Kaspi et al. (2007).  Continuum flux is shown as filled circles,
 and the C IV $\lambda 1550$ line flux as open circles.}
\label{fig7}
\end{figure}
The size of the luminous part of AGN accretion discs $r_{s}$ has been
measured from microlensing in multiply lensed quasar systems by several
groups \cite{p98,k04,m10}.  The results are in good agreement, giving a
value $r_{s} \sim 3 \times 10^{15}$ cm.  The size of the Einstein radius
for a solar mass star from Eq.~\ref{eqn1} is some 10 times larger than
this, which is sufficient for the quasar disc to be microlensed as a point
source \cite{l93}.

The extent to which the broad emission lines in quasar spectra will be
microlensed depends upon the size of the broad line region (BLR).  There
have been extensive efforts to measure the distance of the BLR from the
central source of continuum flux using reverberation mapping techniques,
with the primary motive of measuring the mass of the central black hole.
The results show \cite{k05} that for quasars, the size of the BLR is
typically in excess of 100 light days, or $3 \times 10^{18}$ cm.  This is
100 times larger than the Einstein radius for a solar mass body, resulting
in negligible microlensing amplification of the BLR \cite{r91}.

If microlensing by stellar mass primordial black holes is the cause of
variations in quasar brightness then the results of this Section imply that
changes in continuum flux will not be accompanied by changes in emission
line flux.  This contrasts with the situation where there are intrinsic
changes in continuum flux, in which case the broad emission line flux will
follow changes in the continuum.  To test for this, we shall make use of
spectrophotometric monitoring programes which have measured continuum and
broad line flux in quasar spectra at different epochs.

Fig.~\ref{fig6} shows the change in C IV $\lambda 1550$ flux as a function
of change in continuum flux from a sample of quasars from the Sloan Digital
Sky Survey (SDSS) \cite{w06}.  All quasars from the sample are included
which have S/N $> 8$ for both epochs, and a time lag $\Delta t > 100$
days.  It will be seen that although the continuum varies by over half a
magnitude in both directions, there is no corresponding change in C IV
flux.

A more detailed picture of the way the spectrum changes is given in
Fig.~\ref{fig7}.  This shows continuum and C IV $\lambda 1550$ light
curves for the quasar SBS 1425+606, adapted from Kaspi et al. (2007).
We first note that within each year, on
a timescale of a few weeks, there is some evidence of a weak correlation
between continuum and line flux changes.  This presumably corresponds to
a response of the BLR to changes in continuum strength, as seen in Seyfert
galaxies such as NGC 5548 \cite{p02}.  However, the continuum light curve
is clearly dominated by a rise of 0.6 magnitudes over a period of about 2
years, followed by a gentle decline.  This has no counterpart in the C IV
flux, which remains constant to within $\pm 0.1$ magnitudes.      

\section{The Galactic halo}
\label{sec4}

\begin{figure*}
\centering
\begin{picture} (0,400) (250,0)
\includegraphics[width=1.0\textwidth]{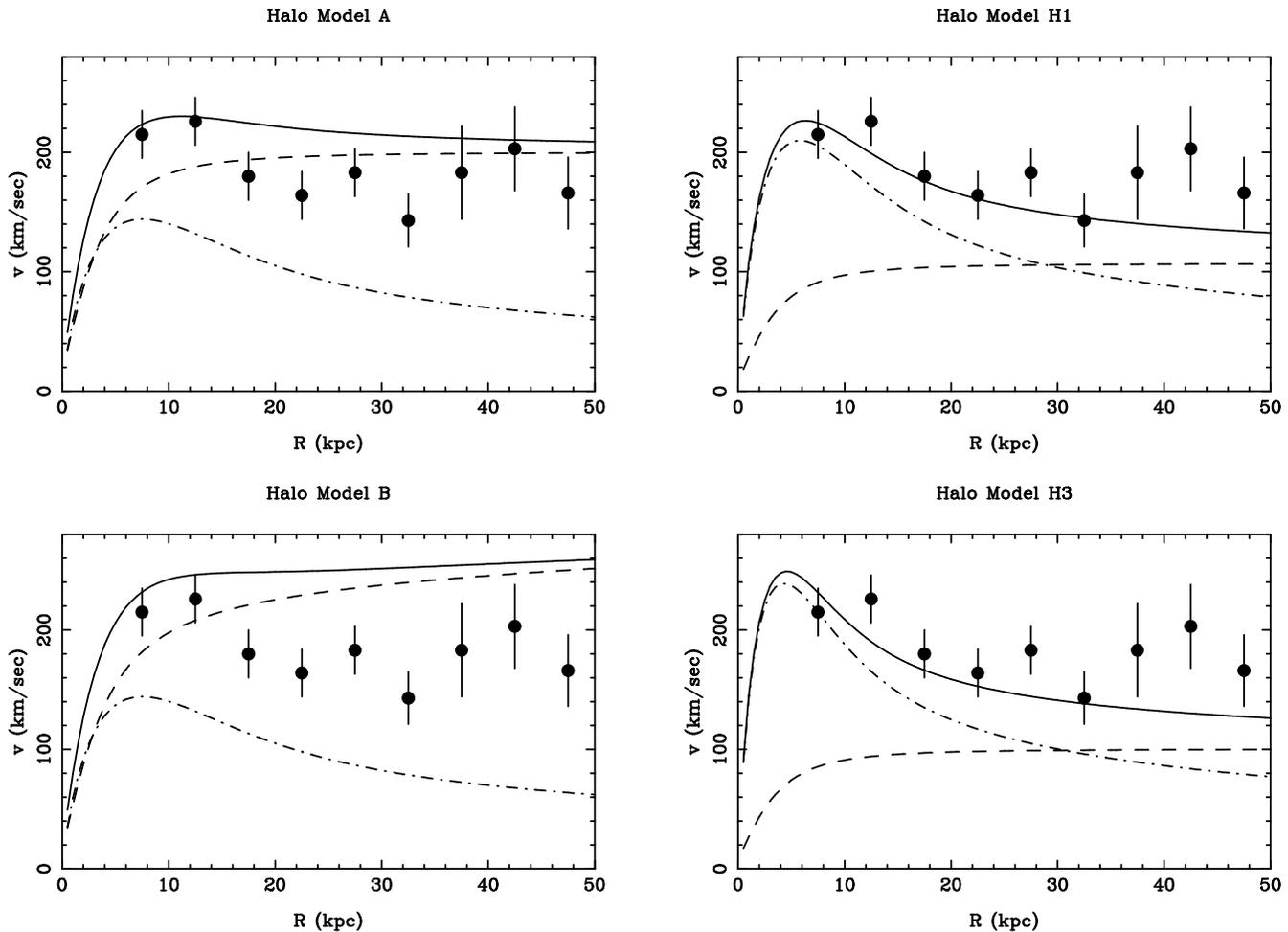}
\end{picture}
\caption{Rotation curves for models of the Galactic disc and halo.  The
 disc and halo components of the rotation curve are shown as dash/dot and
 dashed lines respectively, and the combined rotation curve by a solid line.
 The observed rotation curve derived from SDSS observations of BHB stars
 (Xue et al. 2008) is shown as filled circles.}
\label{fig8}
\end{figure*}

One of the most powerful tests for dark matter in the form of primordial
black holes, or any other compact bodies, is to search for them in the 
halo of the Milky Way.  An ingenious procedure for doing this was proposed
by Paczy\'{n}ski (1986b), which led to the setting up of the MACHO project.
The idea behind this well known project was to photometrically monitor
several million stars in the Magellanic clouds, to search for the rare
event when one of them was microlensed by a compact halo object (MACHO).
This project was challenged by a number of formidable difficulties,
including the creation and running of a suitable photometric pipeline, a
robust algorithm for detecting microlensing events, a reliable estimator
for detection efficiency and an accurate model of the Milky Way halo.
Despite these difficulties, the project yielded excellent results and
three microlensing events were detected in the first year \cite{a96}.
This increased to 17 events when the project was completed some 6 years
later.

Despite the success of the detection of microlensing events, the results
appeared ambiguous.  The team estimated that the detections only accounted
for 20\% of the mass of a typical halo model, and that a halo consisting
entirely of compact bodies could be ruled out at 95\% confidence level for
all but their most extreme halo model.  On the other hand, the population
of compact bodies implied by the microlensing detections was substantially
larger than all known stellar populations \cite{a00a}.  These conclusions
have been sufficient to create a consensus that dark matter is not made up
of compact objects, and to look to elementary particles to provide
alternative candidates.  However, the identity of the of the objects that
were detected still remains unclear.

Since the publication of the MACHO survey results \cite{a00a} there has
been much discussion and disagreement about the reliability of event
classification, the location of the lenses and the estimation of detection
efficiency (Bennett et al. 2005 and references therein).  There are
several phenomena which can mimic microlensing events, including certain
types of variable star and supernovae, and different classification
procedures often do not agree.  The location of the lenses presents an
additional source of uncertainty.  As well as lenses in the Galactic halo
which were the target of the MACHO project, microlensing events can in
principle also be caused by stars in the LMC itself, or in the Galactic
disc.  There seems little doubt that some lensing events have been
mis-classified, but there is still disagreement about how many and which
ones \cite{b05}.  A crucial part of the process for determining the
MACHO content of the halo is the estimation of the detection efficiency
\cite{a00b}.  Not all lensing events that take place during the period
of the survey will be detected, and to convert the detections which are
made to the actual halo population of MACHOs requires a complex computer
simulation.  There are many parameters involved in the detection
probability of an event, including maximum magnification, Einstein radius
crossing time, time of peak magnification, observing programme, weather
conditions and overlaps by neighbouring stars.  In addition the computer
model requires a knowledge of the the stellar luminosity function across
the Magellanic clouds, the distribution of blended images and how this
affects photometric accuracy.  The MACHO collaboration have done a careful
job in modelling these effects, but there are many uncertainties which
could result in either an increase or decrease in the size of the observed
MACHO population of the halo, and no way of making an external check on
the accuracy of the detection efficiency.  With these caveats in mind we
shall adopt the published results of the MACHO collaboration \cite{a00a}
in what follows here.

An important measure of the mass of a population of compact bodies in a
galaxy halo is the optical depth to microlensing $\tau$, defined as
\cite{a96}

\begin{equation}
 \tau = \frac{4 \pi G}{c^2} \int_{0}^{L} \rho(l) \frac{l(L-l)}{L} dl
\label{eqn10}
\end{equation}

\noindent
where $L$ is the distance from observer to source.  This is effectively
the probability that a source is being microlensed by a compact body
along the line of sight.  The calculation of $\tau$ require a
knowledge of the distribution of microlensing mass between the source and
observer, which in the case of the MACHO project means a knowledge of the
mass profile of the Galactic halo.  The MACHO collaboration have made it
clear from the start that their conclusions rely heavily on the chosen
halo model \cite{a95} since, although they can measure quite accurately
the MACHO mass within the distance to the LMC, the total halo mass
fraction in MACHOs is model dependent.  For their analysis they use a
standard model of the form

\begin{equation}
 \rho(r) = \rho_{0} \frac{R_0^2 + R_c^2}{r^2 + R_c^2}
\label{eqn11}
\end{equation}

\noindent
and power law models from Evans (1994), where the density $\rho$ is given by

\begin{eqnarray}
 \lefteqn{\rho(R,z) = \frac{v_a^2 R_c^\beta}{4 \pi G q^2}}
 \nonumber \\
 & & \times \frac{R_c^2(1+2q^2)+R^2(1-\beta q^2)+z^2[2-q^{-2}(1+\beta)]}
     {(R_c^2+R^2+z^2q^{-2})^{(\beta+4)/2}}
\label{eqn12}
\end{eqnarray}

\noindent
where $R$ is the distance from the galactic centre and $z$ is the height
above the plane of the disc.  $q$ is an ellipticity parameter, $\beta$
determines whether the rotation curve is rising or falling, $v_a$ is a
normalisation velocity and $R_c$ is the halo core radius.  Full details
of the model are given by Evans (1994).  The MACHO collaboration galaxy
model also incorporates an exponential thin disc with surface density
$\Sigma$ given by

\begin{equation}
 \Sigma(R) = \Sigma_d \exp{-R/R_d}
\label{eqn13}
\end{equation}

\noindent
where $\Sigma_d$ is a normalization parameter and $R_d$ is the disc scale
length. 

The MACHO collaboration defined a number of model galaxies with values for
$\beta$, $q$, $R_c$ and $R_{d}$.  They also set values for the sun's
galactocentric distance $R_0$, the sun's circular speed $\Theta_0$ and the
local column density $\Sigma_{0}$.  This then sets values for the
normalization parameters $v_{a}$ and $\Sigma_{d}$.  These parameter values
were intended to span the range of currently accepted measurements.  The
collaboration then calculated the optical depth $\tau$ for stars in the
LMC for each halo model, for comparison with the optical depth calculated
from their monitoring programme.  In the first year of observations, the
three detections yielded an optical depth $\tau = 0.88 \times 10^{-7}$
\cite{a96}.  This changed to $\tau = 2.9 \times 10^{-7}$ after the second
year \cite{a97}, and converged to $\tau = 1.2 \times 10^{-7}$ at the
completion of the observations after 5.7 years \cite{a00a}.  In this final
paper the MACHO collaboration chose to restrict their halo models to
three, shown in Table~\ref{tab3} as models S, B and F, spanning the full
range of predicted optical depth to lensing.  On this basis they concluded
that even their most extreme model was excluded, and that a 100\% MACHO
halo was ruled out at a 95\% confidence level.

\begin{table}
\caption{Galactic models for LMC microlensing.}
\label{tab3}
\centering
\vspace{5mm}
\begin{tabular}{l c c c c c c}
\hline\hline
 & & & & & & \\
 Model & S & B & F & H1 & H2 & H3 \\
 & & & & & & \\
\hline
 & & & & & & \\
 $\beta$ & - & -0.2 & 0 & 0 & 0 & 0 \\
 $q$ & - & 1 & 1 & 1 & 1 & 1 \\
 $R_{c}$(kpc) & 5 & 5 & 25 & 5 & 5 & 5 \\
 $R_{0}$(kpc) & 8.5 & 8.5 & 7.9 & 8.5 & 8.5 & 8.0 \\
 $\Sigma_{0}(M_{\odot}$pc$^{-2})$ & 50 & 50 & 80 & 67 & 50 & 50 \\
 $R_{d}$(kpc) & 3.5 & 3.5 & 3.0 & 2.7 & 2.3 & 2.0 \\
 $\Theta_0$(km s$^{-1}$) & 192 & 233 & 218 & 220 & 220 & 226 \\
 $\tau_{LMC}(10^{-7})$ & 4.7 & 8.1 & 1.9 & 1.58 & 1.64 & 1.45 \\
 & & & & & & \\
\hline
\end{tabular}
\end{table}
             
The predicted optical depth to microlensing $\tau$ is sensitive to
most of the parameters listed in Table~\ref{tab3}, and the values
used by the MACHO collaboration reflect the best measurements or
estimates available at the time \cite{a96}.  Since then, there has been
much observational work on the values of these parameters, and there now
seems to be a case for reviewing the MACHO limits on the basis of the new
data.  With this end in view we shall re-work the analysis of the MACHO
collaboration for a new set of models with parameters in line with recent
observational developments.

We first consider the halo core radius $R_{c}$.  This parameter has
conventionally been taken to have a value $R_{c} \sim 5$ kpc, as suggested
by Griest (1991), and supported by the work of Donato et al. (2004).
The values of 20--25 kpc used by Alcock et al. (1996) for all their big
disc models cannot any longer be justified by current observations.  In
fact, providing $R_{c}$ is less than $R_{0}$ it makes little difference to
the value of $\tau$, and so we shall keep $R_{c} = 5$ kpc for our new halo
models.  An important parameter in the Galactic model is the disc scale
length $R_{d}$, for which Alcock et al. (1996) use values of 3--3.5 kpc.
More recent work on the direct measurement of $R_{d}$ gives scale lengths
in the range 2.0--2.7 kpc \cite{o00,s02,j08}, and we shall use values in
this range for our models.  Estimates of the distance to the Galactic
centre $R_{0}$ and the sun's circular speed $\Theta_0$ have not changed
greatly in recent years, and we adopt the same range of values as Alcock
et al. (1996).  Recent work on the local column density $\Sigma_{0}$
\cite{k03} supports the value of 48 $M_{\odot}$ pc$^{-2}$ from Kuijken \&
Gilmore (1991), and there seems to be little observational support for the
large values used by the MACHO collaboration for their big disc models.
We do however include the value of 67 $M_{\odot}$ pc$^{-2}$ from Siebert
et al. (2003) in one of our models. 

The calculation of the optical depth $\tau$ was carried out using the
procedure described by Alcock et al. (1996).  After reproducing their
results, $\tau$ was calulated for a wide variety of disc/halo models.
It was found that with the new range of parameter values discussed above,
there was no difficulty in defining a halo such that the optical depth to
microlensing lay within $1 \sigma$ of
$\tau = 1.2_{-0.3}^{+0.4} \times 10^{-7}$, the value found by Alcock et al.
(2000a) after 5.7 years of observation.  The parameters for three examples
of halos which are not excluded by the MACHO collaboration confidence
limits are given in Table~\ref{tab3} as H1, H2 and H3.

We now consider the sensitivity of the optical depth $\tau$ to changes in
the halo parameters.  This was estimated by re-calculating $\tau$ for the
halo models in Table~\ref{tab3}, but changing each parameter in turn by
1\% and keeping the remainder constant.  The resulting change in $\tau$ is
thus a measure of the sensitivity to that parameter, and is shown in
Table~\ref{tab4} as a percentage, $+$ or $-$ indicating an increase
or decrease in $\tau$ with increase in the parameter.  $\tau$ turns out
to be most sensitive to $R_0$, the distance to the Galactic centre.
However, as dicussed by Olling \& Merrifield (2000), $R_0$ and $\Theta_0$
are not independent, the ratio $\Theta_0/R_0$ being tightly constrained by
measures of the proper motion of SgrA$^*$ \cite{r99}.  This results in
little change in $\tau$ for large changes in $R_0$.  There is little room
for the local column density $\Sigma_0$ to be much less than
50 $M_{\odot}$ pc$^{-2}$, as visible matter is estimated to make up some 
53 $M_{\odot}$ pc$^{-2}$ \cite{h04a}.  Larger values result in a decrease
in the value of $\tau$, implying a lower expected rate for microlensing.
The scale length of the Galactic disc $R_d$ is one of the most uncertain
parameters, and $\tau$ is moderately sensitive to it.  This is because
for given values of $R_0$ and $\Theta_0$, the gravitational effect of a
more massive disc must be compensated for by a less massive halo, leading
to a smaller value for $\tau$.
On the whole, measurements of $R_d$ have tended to become smaller, but
Table~\ref{tab3} reflects the range of currently supported values.  For
the remaining parameters, $\tau$ is relatively insensitive to $R_c$ and
$q$, and changing $\beta$ over the range covered by Alcock et al. (1996)
only results in a small change in $\tau$.

\begin{table}
\caption{Sensitivty of optical depth $\tau$ to halo parameters.}
\label{tab4}
\centering
\vspace{5mm}
\begin{tabular}{r r r r r r}
\hline\hline
 & & & & & \\
 $q$ & $R_c$ & $R_0$ & $\Sigma_0$ & $R_d$ & $\Theta_0$ \\
 & & & & & \\
\hline
 & & & & & \\
 $-0.6$ & $+0.6$ & $-14.6$ & $-4.6$ & $+8.8$ & $+11.0$ \\
 & & & & & \\
\hline
\end{tabular}
\end{table}

Fig.~\ref{fig8} illustrates a selection of the rotation curves from the
halo models in Table~\ref{tab3}.  In each case the solid line shows the
rotation curve for the model, made up of contributions from the halo and
disc.  Also shown is the rotation curve derived from SDSS observations of
some 2400 blue horizontal branch stars by Xue et al. (2008).  Alcock et
al. (1995) point out that each set of parameters should give a model
consistent with the measured Milky Way rotation curve.   The two
left hand panels show rotation curves for massive halo models, which
clearly do not provide an adequate fit to the observed rotation curve.
In this context they consider their Model F to have an extremely low mass
halo, `somewhat inconsistent with the known Galactic rotation curve'
\cite{a00a}.  In fact, Model F has rotation speed at 50 kpc of 160 km
s$^{-1}$, which is close to the value from the more recent rotation curve
of Xue et al.(2008).  The right hand panels of Fig.~\ref{fig8} show two
halo models from Table~\ref{tab3} with optical depth to microlensing
$\tau$ not excluded by the MACHO collaboration limits \cite{a00a}, and
values of $\chi^2$ giving an adequate goodness-of-fit to the observations.
The implication of this is that using more recent measurements of the
structural parameters of the Galactic disc and halo, and the rotation
curve of the Milky Way, the MACHO collaboration results do not exclude a
dark matter halo consisting entirely of compact bodies.
 
\section{Multiply lensed quasars}
\label{sec5}

\begin{figure}
\centering
\begin{picture} (0,200) (120,0)
\includegraphics[width=0.5\textwidth]{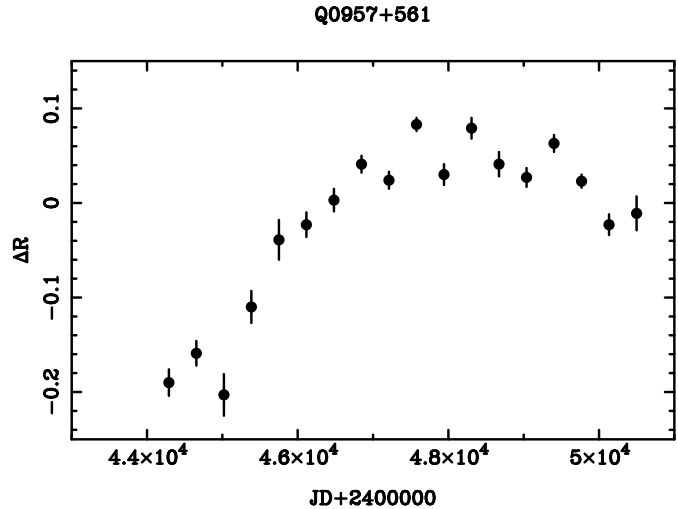}
\end{picture}
\caption{Difference light curve from 1979 to 1997 for the two images of
 the gravitational lens system Q0957+561.  The filled circles are yearly
 weighted means of observations described by Schild \& Thomson (1995)
 and listed at https://www.cfa.harvard.edu/$\sim$rschild/fulldata2.txt}
\label{fig9}
\end{figure}

One situation in which quasar microlensing can be unambiguously
identified is in multiply lensed quasar systems.  These systems
usually arise when a massive galaxy lies close to the line of sight to a
quasar.  The galaxy acts as a gravitational lens, and splits the quasar
light, typically into two or four images.  In these systems it
is well established that the individual quasar images vary partly
synchronously with a short time gap, and partly independently.  The
variations which are seen in both images are taken to be associated with
intrinsic variations of the quasar nucleus, and the time difference
is a measure of the difference in path length to the two images.
However, in all systems which have been studied in sufficient detail
the images also vary independently, and this is generally accepted as
being due to microlensing by compact bodies along the light paths,
where each image would be affected differently.

It is tempting to interpret this microlensing as persuasive evidence
that quasars are being microlensed by a cosmological distribution of
compact dark matter bodies, but there is an important caveat.  By their
very nature, in these quasar systems there will always be a massive
galaxy close to the line of sight, and typically the quasar light
paths will pass through this galaxy.  It has generally been assumed that
the stellar population in the lensing galaxy is sufficient to provide
the population of lensing bodies, but because of the sparse distribution
of stars in galactic halos this assumption may not always be reliable.

The gravitational lens system Q0957+561, comprising two quasar images and
a massive elliptical lensing galaxy, is well suited to investigating the
feasibility of microlensing by stars in the lensing galaxy.  Since its
discovery in 1979 \cite{w79}, the brightness of the two images has been
intensively monitored by several groups.  Schild \& Thomson (1995)
collate many of the earlier observations, which are tabulated at
https://www.cfa.harvard.edu/$\sim$rschild/fulldata2.txt.  Fig.~\ref{fig9}
shows the difference between the $R$ magnitudes of the two images as a
function of time, covering the period from 1979 to 1998.  Yearly weighted
means were taken for the two images $A$ and $B$, using the observational
errors given with the data.  The difference in magnitude $m$ was then
plotted in the sense $\Delta R = m_A-m_B$, with error bars showing the
dispersion of the measures within each yearly epoch.

The variation of around 0.3 mag seen in Fig.~\ref{fig9} is generally
accepted as being due to microlensing, as no other plausible mechanisms
have been suggested.  It is also a lower limit on the total microlensing
amplitude, since when both images are brightening or fading, it
underestimates the variation in each image separately.  We now address
the question of whether this microlensing is likely to be due to stars
in the lensing galaxy.

The lensing galaxy for the Q0957+561 system is a giant elliptical lying
close to the line joining the two images.  For Hubble constant $H_0 = 70$
km sec$^{-1}$ Mpc$^{-1}$, the distances from the centre
of the galaxy to images A and B are 41.9 kpc and 8.9 kpc respectively.
The photometry and morphology of this galaxy have been studied in some
detail by Bernstein et al. (1997) using WFPC2 observations from the HST.
They fit elliptical profiles to the galaxy image which gives surface
brightness values of 25.2 and 22.6 $V$ mag arcsec$^{-2}$ at the positions
of the A and B images respectively.  We convert these surface
brightnesses to mass surface densities $\Sigma$ using a mass-to-light
ratio for old stellar populations derived from globular clusters
\cite{m05}.  Adopting $M_{\odot} / L_{\odot} = 2$ as a representative
value gives $\Sigma_A = 5.2$ $M_{\odot}/L_{\odot}$ pc$^{-2}$ and
$\Sigma_B = 56.9$ $M_{\odot}/L_{\odot}$ pc$^{-2}$.

To calculate the optical depth to lensing $\tau_A$ and $\tau_B$ for
images A and B we use a modified version of Eq.~\ref{eqn10}:

\begin{equation}
 \tau = \frac{4 \pi G}{c^2} \frac{\Sigma D_g(D_q-D_g)}{D_q}
\label{eqn14}
\end{equation}

\noindent
where $D_g$ and $D_q$ are the distances to the lensing galaxy and quasar
respectively.  Using the values for $\Sigma_A$ and $\Sigma_B$ derived
above we obtain $\tau_A = 4.0 \times 10^{-3}$ and
$\tau_B = 4.4 \times 10^{-2}$.  These low optical depths to lensing
imply that a strong microlensing feature of the type illustrated in
Fig.~\ref{fig9} is very unlikely to be produced by stars in the lensing
galaxy.  This leaves two other possibilities.  If the dark matter in the
lensing galaxy were in the form of compact bodies, then the probability
of microlensing would be much larger.  Alternatively, a cosmological
distribution of dark matter in the form of compact bodies could be
responsible for the observed microlensing.  Either way, this result
implies dark matter in the form of stellar mass compact bodies. 

\section{Candidates for compact dark matter}
\label{sec6}

Hitherto in this paper we have summarised the evidence supporting the
idea that dark matter is predominantly in the form of compact bodies,
the most plausible candidates being primordial black holes.  In this
section we shall consider the various possibilities in more detail.

\subsection{Baryonic bodies}

Perhaps the most plausible candidates for compact dark matter bodies are
low mass stars and stellar remnants.  For stars with mass 
$M > 0.1 M_{\odot}$ the observed background radiation density puts tight
constraints on $\Omega_{\ast}$, the stellar contribution to dark matter,
and on this basis Carr (1994) finds $\Omega_{\ast} < 0.1$.  For stars
with $M < 0.1 M_{\odot}$ Gilmore (1999) uses arguments for the
universality of the IMF and direct mid-infrared observations to constrain
the contribution of $\Omega_{\ast}$ in galactic halos to negligible
proportions.  Notwithstanding observations such as these, there would
always remain the possibility that baryonic matter could remain hidden in
some unexpected form were it not for the constraints of cosmological
nucleosynthesis, which we now consider.

The standard theory of primordial nucleosynthesis makes predictions for
the abundance of the light elements D, $^3$He, $^4$He and $^7$Li.  These
abundances depend, with varying sensitivity, on the baryon to photon ratio
$\eta$.  Comparison of the predictions with astronomical observations of
the abundances only shows consistent agreement over a narrow range of
$\eta$.  This consistency has the dual effect of giving strong support
to the standard theory of primordial nucleosynthesis, and providing a
robust measure of $\eta$.  The photon density may be readily obtained
from measures of the Cosmic Microwave Background, giving a value for
the baryon density $\Omega_b$ in the range
$0.011 h^{-2} < \Omega_b < 0.15 h^{-2}$ \cite{s93}.
For $H_0 = 70$ km sec$^{-1}$ Mpc$^{-1}$ this gives $\Omega_b < 0.03$.
This limit effectively rules out any candidates for dark matter in the
form of compact bodies which are made up of baryons, and certainly
excludes familiar bodies such as stars and planets and any remnants of
their evolution.

\subsection{Primordial black holes}

The only non-baryonic compact bodies which have so far been seriously
proposed as dark matter candidates are primordial black holes \cite{c94}.
Hawking (1971) showed that density fluctuations in the early Universe
could lead to gravitational collapse, and Carr \& Hawking (1974) argued
that this collapse could lead to the formation of black holes of around
the horizon mass $M_h$ at time $t$, where

\begin{equation}
 M_h(t) \approx \frac{c^3 t}{G} 
\label{eqn15}
\end{equation}

\noindent
Carr (1975) examined the expected primordial black hole mass spectrum,
and concluded that the density fluctuations invoked to seed the formation
of galaxies are of a type to favour primordial black hole production over
a large mass range.  As density fluctuations larger than the horizon size
enter into the particle horizon they can be close to their Schwarzschild
radius, and depending on the balance between pressure and gravity may
collapse to become black holes.  For a hard equation of state where
pressure $p$ and energy density $\rho$ are equated by $p = \rho / 3$,
regions with overdensity $\delta \rho / \rho \gtrsim 1/3$ should collapse
into black holes of the order of the horizon mass $M_h$ from
Eq.~\ref{eqn15}.  Fluctuations smaller than this will tend to disperse
due to pressure forces.  For a soft equation of state where $p \approx 0$,
black holes should form much more readily from small relic adiabatic
perturbations.  The problem with this approach as it stands is that a
scale invariant power spectrum normalised to the CMB fluctuations
produces a negligible density of primordial black holes.

\begin{figure*}
\centering
\begin{picture} (0,200) (250,0)
\includegraphics[width=1.0\textwidth]{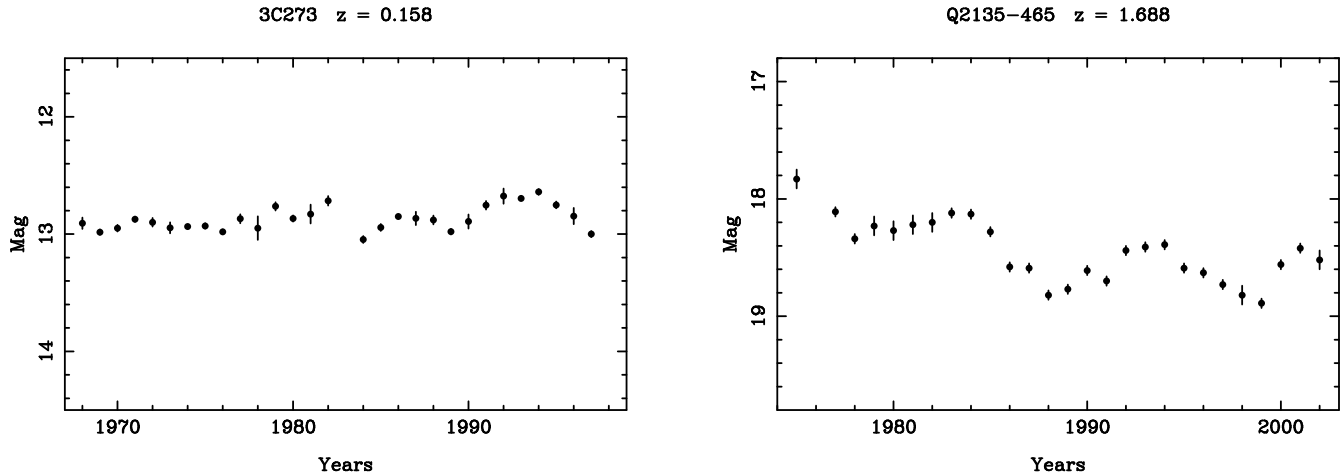}
\end{picture}
\caption{The left hand panel shows the $V$ band light curve for the
 nearby quasar 3C273 from the 3C273 database at
 http://isdc.unige.ch/3c273.  The right hand panel shows the $B_J$-band
 light curve for a high-redshift quasar from the Field 287 survey.}
\label{fig10}
\end{figure*}

During phase transitions in the early Universe, and in particular during
the quark-hadron transition of the QCD epoch at a cosmic epoch of
$\sim 10^{-5}$ sec, the equation of state is expected to soften.  Jedamzik
(1997) concluded that the primordial black hole mass function should show
a pronounced peak at the QCD mass scale $M_{QCD} \approx 1 \: M_\odot$.
This result was confirmed by general relativistic hydrodynamic numerical
simulations which showed that the fluctuation density threshold 
$\delta \rho / \rho$ for black hole formation falls when the Universe
undergoes phase transitions, to produce a mass spectrum dominated by the
horizon masses at the transition epochs \cite{j99}.  Bullock \& Primack
(1997) argued that non-Gaussian density fluctuations produced during
inflation could prevent over-production of black holes, as well as
providing a source of large amplitude fluctuations.  This would be
consistent with black hole production at the QCD mass scale, although
Widerin \& Schmid (1998) claimed that some additional fine tuning would
be necessary.  Yokoyama (1997) adopted a different approach, making the
case that for inflation models with multiple scalar fields, isocurvature
fluctuations are generated which, with a suitable choice of model
parameters, produce primordial black holes with masses peaked at the MACHO
mass scale.  This idea has been developed \cite{k08a,f10} in the context
of the double inflation models.  In this framework, primordial black holes
with a narrow mass distribution can be produced in sufficient abundance
to make up the dark matter, and with natural values for the model
parameters will produce masses in the MACHO range.

\section{Discussion}
\label{sec7}

\begin{figure*}
\centering
\begin{picture} (0,280) (250,0)
\includegraphics[width=1.0\textwidth]{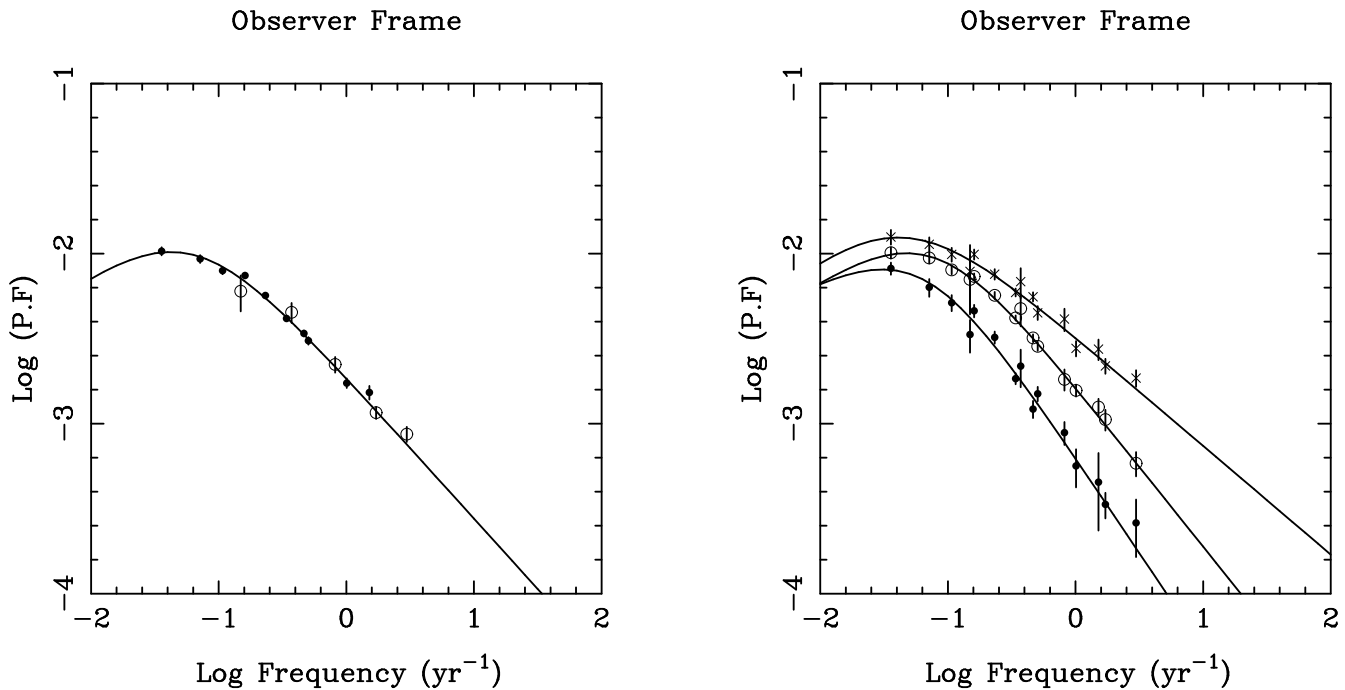}
\end{picture}
\caption{The left hand panel shows SEDs of light curves in the
 observer frame for quasars from the Field 287 survey (filled
 circles) and from the MACHO project (open circles).  The solid line is
 the best fit of the curve in Eq.~\ref{eqn7}.  The right hand panel 
 shows the same data divided into three magnitude ranges.  In this case
 filled circles, open circles and stars represent high, medium and low
 luminosity bins respectively.  Solid lines are fits to the data as for
 the left hand panel} 
\label{fig11}
\end{figure*}

For the last 20 years or so there has been a generally held view
that dark matter is predominantly in the form of elementary particles.
This is based in part upon the belief that there are no other suitable
non-baryonic candidates, but it is also motivated by the limits on MACHOs
derived from microlensing studies.  This paper has set out to make the
case that stellar mass primordial black holes provide viable alternative
candidates.  These bodies are not sufficiently massive to form luminous
accretion discs, and can only be detected through the effects of
gravitation, in particular gravitational lensing.  In the case of bodies
of around a solar mass, the changes in the lensing pattern have a
timescale of the order of several years, which opens up the possibility
of detecting them by observations of microlensing. 

If primordial black holes make up a substantial proportion of dark matter,
then every line of sight will be distorted at some level by gravitational
lensing \cite{p73}.  This means that for compact light sources such as
quasars, the effects of microlensing should be detectable in their light
curves.  Variations due to microlensing exhibit a number of well-defined
characteristics which, if observed, would thus indicate the presence of
primordial black holes.  In Section~\ref{sec2} we focus on four
properties of the light curves that should be present if the variations
in flux are predominantly the result of microlensing.  Ideally, one would
like to compare the predictions for microlensing and for 
models of intrinsic variation such as accretion disc instability, with
observed quasar light curves.  Unfortunately, at present no models of
intrinsic variation make predictions that are sufficiently precise
to be useful for comparison with observations.  Microlensing on the
other hand results in specific predictions which can be tested
experimentally.

There is one test which can distinguish between microlensing and intrinsic
variations, in spite of the difficulties of modelling the latter.
The absence of time dilation seen in Fig.~\ref{fig2} is not
consistent with any mechanism for intrinsic variabilty in quasars, but it
is in agreement with the predictions for microlensing.  Of the other three
properties of quasar light curves described in Section~\ref{sec2}, none
can strictly speaking exclude intrinsic variation, given the lack of
constraint on models of quasar emission.  However, the purpose of this
paper is to show that the observations are consistent with the much more
precise predictions for microlensing by primordial black holes.
Microlensing simulations allow one to predict the shape
of the light curve SEDs, and as can be seen from Fig.~\ref{fig1} there is
good agreement with the observations.  Without this agreement the case
for microlensing would be severely weakened.

The question of colour changes with variation in flux is more complicated.
It is well known that gravitational lensing is essentially an achromatic
process \cite{s92}, with the expectation that for microlensing there will
no change in colour with brightness.  However, in the case of a resolved
source, that is a source significantly larger than the Einstein radius of
the lenses, this will not necessarily be true.  For example, if the source
has a radial colour gradient, blue in the centre and reddening outwards,
then the blue light profile, being more compact, will be preferentially
amplified relative to the red as the lenses cross the source \cite{r91}.
In fact, as we have seen in Section~\ref{sec3}, for bodies around a solar
mass the Einstein radius is typically much larger than the quasar
accretion disc and so the variations are expected to be achromatic.  This
is what was found for the quasar light curves analysed in
Section~\ref{sec2}, in agreement with the predictions for primordial
black holes.  Another property predicted for light curves from
microlensing is statistical symmetry, and again the results in
Section~\ref{sec2} show no evidence for asymmetry.  It is not clear to
what extent predictions for intrinsic variations favour symmetry.  Despite
the work of Kawaguchi et al. (1998), they are most likely to be model and
timescale dependent.  The point of significance for this paper is that
all the features of quasar light curves analysed here are consistent with
the predictions for microlensing by solar mass primordial black holes.

Changes in quasar spectra with variations in brightness provide an
alternative way of testing for the presence of compact bodies.  There is
abundant evidence that broad emission lines respond to changes in
continuum brightness, but the presence of a near critical density of
primordial black holes implies that the continuum source will also vary
with no corresponding change in broad line strength as a result of
microlensing.  This is clearly seen in Figs~\ref{fig6} and \ref{fig7},
and the evidence points to longer timescales of variation being associated
with microlensing events.  In the case of multiple quasar systems, the
intrinsic features seen in both light curves, and used to measure time
delay, are typically of small amplitude and duration.  On the other
hand, microlensing events as measured from the difference between the
light curves of individual images tend to be of large amplitude and to
last for tens of years, as seen in Fig.~\ref{fig9}.

The various constraints discussed above suggest a picture in which for low
luminosity Seyfert I galaxies, observed variations are dominated by short
timescale changes in flux accompanied by changes in colour and broad
emission line strength.  For more luminous quasars the amplitude of these
intrinsic variations decreases, and at higher redshift the light curves
become dominated by the effects of microlensing, as illustrated in
Fig.~\ref{fig10}.  These variations occur on a longer timescale of around
20 years, are close to being achromatic, and do not affect the broad line
region.  The expectation from Eq.~\ref{eqn9} is that with increasing
luminosity, the amplitude of variation will become smaller as the size of
the emitting region gets larger.

The idea that there is a correlation between the way quasars vary and
their absolute magnitude or luminosity has a long history.  In
particular, several authors \cite {h94,c96,h00} have claimed to find
an anti-correlation between luminosity and amplitude, in the sense that
for a sample of quasar light curves, more luminous quasars are seen to
vary over a smaller range of brightness than less luminous ones.  One
of the problems with this conclusion is that the observed amplitude is
clearly a function of the length of the run of observations, and so
can be confused with timescale of variability.  Fourier analysis
provides a way round this by giving measures of variability on different
timescales.  The left hand panel of Fig.~\ref{fig11} shows the combined
SED for the samples of light curves described in Section~\ref{sec2}, and
fitted with the function $P(f)$ in Eq.~\ref{eqn7}.  In the right hand
panel the data is divided into three lumnosity ranges, $M_B > -23.5$,
$-25.5 < M_B < -23.5$ and $M_B < -25.5$.  The three curves show
broadly the same features as the curve in the left hand panel, but it
is clear that the anti-correlation between luminosity and amplitude is
confirmed.  The maximum power density of the lowest luminosity quasars
is greater than the highest by a factor of 1.5.  In this case it
appears that the time span of the data is sufficient to resolve the
degeneracy between timescale and amplitude.  This is consistent with the
model outlined above.

For some time here has been a tendency to dismiss the idea that dark
matter is in the form of primordial black holes or other compact bodies
on the grounds that the MACHO collaboration observations of LMC stars have
conflicted with predictions of the microlensing event rate.  This would
mean the falsification of the theory, regardless of the arguments in its
favour \cite{p59}.  We have shown in Section~\ref{sec4} that using more
recent parameters for Galactic structure and dynamics there need be no
discrepancy between predicted and observed event rates, and hence there
are viable halo models where the dark matter is in the form of stellar
mass compact bodies.  The results of  Section~\ref{sec5} even suggest
that such bodies may have already been routinely detected in photometric
monitoring of multiply lensed quasar systems.  This means that the case
for primordial black holes as dark matter can be considered on its merits,
and contrasted with arguments favouring dark matter in the form of
elementary particles. 

\section{Conclusions}
\label{sec8}

The main conclusion of this paper is that stellar mass primordial black
holes are plausible dark matter candidates.  These objects would have
been created during the QCD phase transition at a cosmic epoch of
around $10^{-5}$ sec.  Such bodies are not easy to detect, but should
betray their presence through the effect of their gravitational fields.
The most readily observable manifestion of this is the gravitational
microlensing of quasars.

We have used power Fourier spectrum analysis of a large sample of quasar
light curves to compare with the results of numerical simulations, and
the predictions for time dilation, colour change and statistical symmetry
from microlensing by a near critical density of primordial black holes.
We have also compared the predictions for spectral changes as a result
of microlensing with spectrophotometric observations of quasars.  Finally,
we make the case that the observed microlensing in multiple quasar
systems is best explained by a population of dark matter bodies rather
than stars. 

The main objection to the idea of dark matter in the form of compact
bodies has come from the search for MACHOs in the Galactic halo.  By
observing the microlensing rate of stars in the Magellanic Clouds, and
adopting a dark matter model for the halo, the MACHO collaboration deduced
a mass fraction in compact bodies of around 20\%.  We have re-worked
their analysis using using the most recent values for the structure and
dynamics of the halo, and find that there is now no conflict between the
observed microlensing rate and a MACHO dominated halo.

The results of this paper do not establish that dark matter is made up
of stellar mass primordial black holes.  However, they do show that if
it is in that form, then a wide variety expected signatures are confirmed
by observation.  On this basis, primordial black holes should be seen as
viable alternatives to elementary particles in the search for the
identification of dark matter.

\section*{Acknowledgements}

This paper is based in part on SuperCOSMOS measurements of UK Schmidt
Telescope plates.  Data are available online through the SuperCOSMOS
Science Archive: http://surveys.roe.ac.uk/ssa

\label{lastpage}
\end{document}